\documentclass{article}
\usepackage[preprint]{neurips_2020}
\usepackage{geometry}
\usepackage{graphicx}
\usepackage{qtree}
\usepackage{booktabs}
\usepackage{siunitx}
\usepackage{ragged2e}
\usepackage{rotating}
\usepackage{adjustbox}
\usepackage{caption}
\usepackage[T1]{fontenc}
\usepackage{pdflscape}
\usepackage{nameref}
\usepackage{pdfpages}
\usepackage[multiple]{footmisc}
\usepackage{float}
\newcolumntype{R}[1]{>{\RaggedRight}p{#1}}
\usepackage{amssymb}
\usepackage{xcolor}
\usepackage{verbatim}
\usepackage{subcaption}
\usepackage{array}
\usepackage{natbib}
\usepackage{hyperref}
\usepackage{mathtools}
\usepackage{xspace}
\usepackage{cleveref}
\usepackage{svg}
\usepackage{transparent}
\usepackage[toc,title,page]{appendix}
\definecolor{orangefull}{RGB}{230, 159, 0}
\newcommand*{\eg}{e.g.\@\xspace}
\usepackage{siunitx}

\colorlet{orange}{orangefull!20!white}
\definecolor{bluefull}{RGB}{86, 180, 233}
\colorlet{blue}{bluefull!20!white}
\definecolor{green}{RGB}{0, 158, 115}
\colorlet{green}{green!20!white}

\title{Climate data selection for multi-decadal wind power forecasts}
\author{%
  Sofia Morelli \thanks{equal contribution}\\
  University of Tübingen\\
  \texttt{sofia.morelli@uni-tuebingen.de} \\
  \And
  Nina Effenberger \footnotemark[1] \\
  Cluster of Excellence Machine Learning\\
  University of Tübingen\\
  \texttt{nina.effenberger@uni-tuebingen.de} \\
  \And
  Luca Schmidt \footnotemark[1]  \\
  Cluster of Excellence Machine Learning\\
  University of Tübingen\\
  \texttt{luca.schmidt@uni-tuebingen.de} \\
  \And
  Nicole Ludwig \\
  Cluster of Excellence Machine Learning\\
  University of Tübingen\\
  \texttt{nicole.ludwig@uni-tuebingen.de}}
  
\begin{document}

\maketitle
\begin{abstract}
    Reliable wind speed data is crucial for applications such as estimating local (future) wind power. Global Climate Models (GCMs) and Regional Climate Models (RCMs) provide forecasts over multi-decadal periods. However, their outputs vary substantially, and higher-resolution models come with increased computational demands. In this study, we analyze how the spatial resolution of different GCMs and RCMs affects the reliability of simulated wind speeds and wind power, using ERA5 data as a reference. 
    We present a systematic procedure for model evaluation for wind resource assessment as a downstream task. Our results show that higher-resolution GCMs and RCMs do not necessarily preserve wind speeds more accurately. Instead, the choice of model, both for GCMs and RCMs, is more important than the resolution or GCM boundary conditions. The \textit{IPSL} model preserves the wind speed distribution particularly well in Europe, producing the most accurate wind power forecasts relative to ERA5 data. 
    %A diverse selection of RCMs may provide a useful spread of simulations for reliable uncertainty quantification. 
    %The analysis shows that GCMs are useful for wind power forecasting, and RCMs add further value, emphasizing the need for high-quality models at both global and regional scales.

\end{abstract}

\section{Introduction}
\label{introduction}
%% WIND FROM CLIMATE MODELS
% wind is an imporant soruce of power and we need long-term forecasts with climate models
Wind is expected to become a dominant source of energy in future power supply \citep{iea}. For wind to be a reliable source of energy, accurate regional wind forecasts over multi-decadal periods are necessary. Climate model output is a natural choice of data for such analyses, as they provide projections on multi-decadal timescales \citep[e.g.][]{pryor2020climate, ndiaye2022projected}. 
%the climate model choice is important: This is very natural
Climate models are complex physical earth system models whose outputs can differ substantially with large inherent model uncertainties \citep{zhang2024quantify}. 
Given this variability, it is crucial to assess the skill of climate models, especially their ability to perform in specific applications of interest such as wind power forecasting \citep{moemken2016decadal, isphording2024standardized}.

%%RCMs and GCMs and downscaling
%however, it is unclear which spatial resolution we need, global vs regional (CMIP5+6), downscaling
Previous research revealed that a higher spatial wind data resolution is more desirable. Not only because it provides more local forecasts \cite{rummukainen2016added}, but because the resolution at which a model operates affects which weather phenomena it can resolve \citep{lucas2021convection}. While higher-resolution models are able to resolve more fine-grained weather phenomena, they come with computational challenges. All models --no matter the spatial resolution-- require modeling assumptions and therefore contain errors and uncertainties. It remains unclear which climate data, at which spatial resolution, best represents future wind speeds.

Addressing this question requires a careful evaluation of the different climate model data available. CMIP6 represents the most recent effort to organize and standardize global climate models (GCMs) \citep{eyring2016overview} and making the resulting multi-model output publicly available. Compared to its predecessor CMIP5, CMIP6 offers significant improvements in model performance \citep{carvalho2021wind}, particularly regarding surface wind speed simulations \citep{miao2023evaluation}. A key limitation of GCMs remains their low spatial resolution, typically at $\geq 100 \text{ km}$. 

There are techniques to increase the spatial resolution of global climate model output. Regional climate models use GCMs as their input and run at finer resolution \citep{vautard2021evaluation} to resolve more fine-grained weather phenomena physically. This process, called dynamical downscaling, is, however, computationally intensive. Statistical downscaling methods offer a less resource-demanding alternative, establishing relationships between low-resolution GCM output and localized weather patterns \citep[\eg][]{schmidt2024wind}. However, these methods still require high-resolution data to establish such relationship. 
%before regional climate models (RCMs) use their output and run at a higher resolution \citep{vautard2021evaluation}. 
%While dynamical downscaling uses high-resolution data from RCMs to refine GCM output, it is computationally intensive \cite{vautard2021evaluation}. 
While RCMs and statistical downscaling techniques can add value in regions with complex terrain or significant land-sea contrasts, they do not always outperform raw GCM data \citep{rummukainen2016added}. Furthermore, for CMIP6 there is no unified collection of RCM data available. Given these data limitations, there is a need to investigate whether raw GCM data from CMIP6 can still be valuable for multi-decadal wind power forecasts–despite their lower spatial resolution. 

The value of climate models is difficult to assess as climate model outputs are not spatio-temporally aligned. Comparing climate model output to reanalyses such as ERA5 is, therefore, non-trivial. Typically, climate models are validated based on aggregate statistics using their historical runs \citep[e.g.][]{randall2007climate}. 
%% SPEED TO POWER
In the context of wind power evaluation, an additional challenge arises as the actual variable of interest--wind power--is a derived variable. The variable projected by climate models is wind speed and the relationship between power and speed is complex. To still account for the known non-linearities, other research considers wind speeds cubed \citep{miao2023evaluation}, simulates load factors \citep{macleod2018transforming} or transforms wind speeds to wind power using a turbine power curve \citep{ko2022projected}.

In the following, we investigate historical wind speeds and derived wind power to analyze how the spatial resolution of global and regional climate model data affects multi-decadal wind power forecasts. In \Cref{methods} we explain how we evaluate different wind (power) projections, in \Cref{results} we describe our results and in \Cref{discussion} we discuss our results as well as shortcomings and future work.

\begin{comment}
    
For such forecasts 
\begin{itemize}
    \item Motivation for wind power in Europe, why multi-decadal
    \item for that we can use climate model data, some references of other research
    \item CMIP6 vs CMIP5 (regional vs global)
    \subitem What are the advantages of higher resolution: They can resolve other weather phenomena
    \subitem Disadvantages: Very expensive, non-existent yet
    \item wind speed vs power -> due to the non-linearities we need distributions and extremes
    \item ERA5
    \item even though climate models give us time series they can not be interpreted as time series, we have to evaluate them differently: cumulative wind power. Assumption: Time of power generation does not matter, we would need batteries. 
    \item What about ensemble average vs single members? We show that averages are less representative. 
\end{itemize}
\end{comment}

\section{Methods}
\label{methods}
To compare the influence of climate model resolution on multi-decadal wind power forecasts, we investigate wind speeds using climate models and subsequent wind power predictions. In the following, we describe the data we use and how we perform evaluations.

\subsection{Data}
\begin{figure}
    \centering  \includegraphics[width=\textwidth]{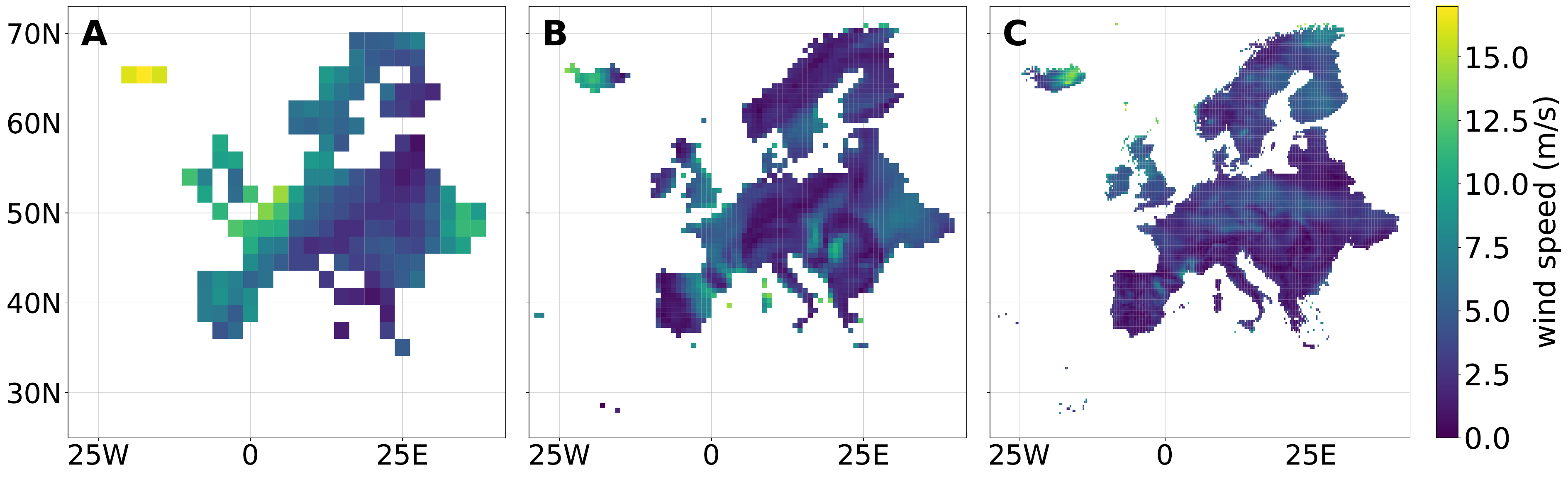}
    \caption{Example of gridded wind speed data over Europe. A single time frame of the same day (01.01.2005) of the lowest resolution CMIP6 data set NCC-LR (A), the highest resolution CMIP6 data set MOHC-HR (B) and the ERA5 reanalysis data set (C).}
    \label{fig:grid-europe}
\end{figure}

In our analysis, we compare the reanalysis dataset ERA5 \citep{hersbach2020era5} to global climate model data from ten CMIP6 models \citep{eyring2016overview} as well as regional climate model data from CORDEX \citep{giorgi2015regional} based on CMIP5 global boundary models \citep{taylor2012cmip5overview}. An overview of all GCMs and RCMs used can be found in \Cref{tab:models_information} and \Cref{tab:models_information_rcms}, respectively. We use wind velocities $u$ and $v$ at six-hourly instantaneous resolution as these were found to be reliable for multi-decadal predictions in \citet{effenberger2023mind}) and compute wind speeds $w$ as 

\begin{equation}
    w = \sqrt{v^2+u^2}.
\end{equation}
Our study region covers all data points in continental Europe within a rectangle with longitudes $\in (25^{\circ}, 73^{\circ})$ and latitudes $\in (-30^{\circ}, 42^{\circ})$. This means we consider gridded data over land, as shown in \Cref{fig:grid-europe}. We focus on historical runs of climate models from 2005 to 2015. We choose the first ensemble member run typically labeled as \emph{r1i1p1f1}. %\textcolor{red}{Sofias Masterarbeit zitieren als Motivation für member statt mean}  
The original wind speeds $w_{10}$ are available at \SI{10}{\meter} above ground\footnote{For the MPI CMIP5, the only available six-hourly wind speed data at the time of research, provided on the ESGF node \citep{MPI_CMIP5_historical}, were at an altitude of \SI{1500}{\meter}. Therefore, $w_{10}$ changes to $w_{1500}$ in the power law to extrapolate the CMIP5 GCM model data to \SI{126}{\meter}.}. To account for common wind turbine hub heights, we interpolate all wind speeds to \SI{126}{\meter} using a wind profile power law following \begin{equation}
    w_{126}=\left(\frac{126}{10}\right)^\alpha \cdot w_{10},
\end{equation}
where the coefficient $\alpha$ is empirically derived to be about $\frac{1}{7}$ for neutral stability conditions \citep{peterson1978use}. 
%For the MPI CMIP5, the only available six-hourly wind speed data at the time of research, provided on the ESGF node \citep{MPI_CMIP5_historical}, were at an altitude of \SI{1500}{\meter}. We, therefore, apply the power law to extrapolate the CMIP5 GCM model data to \SI{126}{\meter}. Additionally, 
Since the regional model runs from CMIP5 are on a curvilinear grid,  we re-grid them to a regular Gaussian grid with a matching spatial resolution (\SI{0.1}{\degree}) using bilinear interpolation \citep{schulzweida_2023_10020800}.
\subsection{Evaluation and metrics}
The evaluation is performed on two aspects: wind speed and wind power. While wind power is our ultimate variable of interest, the analysis primarily centers around wind speed, the variable provided by the climate models. Wind power can be estimated from wind speed using a theoretical wind power curve. Due to the non-linear nature of this relationship, however, a comprehensive analysis of wind speed distributions is essential for understanding its implications for wind power.

\subsubsection{Wind speed distributions}
%We begin by examining the basic statistical moments of wind speed data, specifically the mean and the variance. While not our primary focus, these quantities provide a basic understanding of the main tendencies and variability of wind speed projections by the different models.
Due to the non-linear relationship between wind speed and wind power ({compare \Cref{fig:power-curve}), models must capture the entire wind speed distribution and not only the mean quantities.
To analyze these distributions, we investigate kernel density estimations (KDEs) and use quantitative methods to compare wind speed distributions. Since there is no universal metric for comparing distributions, we employ the Jensen-Shannon distance and the Wasserstein-1 distance, each providing a distinct approach to quantifying differences between distributions.

\begin{comment}
\paragraph{QQ-Plots}
QQ-Quantile-Quantile  plots are used to compare the different quantiles of the wind speed distributions from climate models against the ERA5 reference distribution.
Using QQ plots we can identify regions that are misrepresented by the respective climate models and build the basis for further investigation.     
\end{comment}
imates}
We estimate dind speed probionbility distributions using KDEskernel density estimations ()

\begin{equation}
    \text{KDE}(w) = \frac{1}{nh} \sum_{i=1}^{n} K\left(\frac{w - w_{i}}{h}\right) \text{\,,}
\end{equation}
using Scott's rule \cite{scott2010scott} to select a bandwidth $h$ and the Gaussian kernel $K(x)$ 

\begin{equation}
    K(x) = \frac{1}{\sqrt{2\pi}} \exp\left(-\frac{1}{2}x^2\right) \text{\,.}
\end{equation}

\paragraph{Jensen-Shannon distance}
We use the Jensen-Shannon (JS) distance to measure how closely a GCM's KDE wind speed distribution aligns with the ERA5 KDE wind speed distribution. This metric is a symmetrical and smooth variant of the Kullback-Leibler (KL) divergence \citep{endres2003new}, and is defined as

\begin{equation}
    \text{JS}(P_{\text{ref}}, P_{\text{pred}}) = \sqrt{\frac{1}{2} \text{KL}(P_{\text{ref}} \parallel M) + \frac{1}{2} \text{KL}(P_{\text{pred}} \parallel M)} \text{\,,}
\end{equation}
where $M$ is the mixture distribution 
\begin{equation}
 \text{M} = \frac{1}{2}(P_{\text{ref}} + P_{\text{pred}}) \text{\,,}
\end{equation}
% Add that value of 1 means most dissimilar
and $ \text{KL}(P_{\text{ref}} \parallel P_{\text{pred}})$ is the KL divergence between the reference distribution $P_{\text{ref}}$ and the predicted distribution $P_{\text{pred}}$, defined as

\begin{equation}
    \text{KL}(P_{\text{ref}}\parallel P_{\text{pred}})=\sum_{x \in \mathcal{X}} (P_{\text{ref}}(x) \log \left(\frac{P_{\text{ref}}(x)}{P_\text{{pred}}(x)}\right) \text{\,.}
\end{equation}
An advantage of the JS distance is that there are no issues related to undefined or infinite values when comparing distributions with zero wind speeds. Additionally, the JS distance is bounded between $[0, 1]$ (unlike the Wasserstein-1 distance), offering an intuitive interpretation of the results as a similarity measure. 

\paragraph{Wasserstein-1 distance}
As an additional way of comparing the wind speed distributions between the GCMs and ERA5 we use the Wasserstein-1 (W1) distance. The metric is based on the concept of \emph{optimal transport} and differs from JS distance in interpretation and mathematical properties. The W1 distance is defined as

\begin{equation}
    \mathrm{W}(P_{\text{pred}}, P_{\text{ref}}) = \inf_{\gamma \in \Pi(P_{\text{pred}}, P_{\text{ref}})} \mathbb{E}_{(x,y) \sim \gamma} [ \| x - y \|_1 ] \text{\,,}
\end{equation}

where $\Pi(P_{\text{pred}}, P_{\text{ref}})$ is the set of couplings, i.e., joint distributions whose marginals are $P_{\text{pred}}$ and $P_{\text{ref}}$. Conceptually, the W1 distance quantifies the expected cost of the optimal transport plan that transforms the distribution $P_{\text{pred}}$ into $P_{\text{ref}}$ \citep{arjovsky2017wasserstein}.

\paragraph{Extreme wind speeds}
Finally, we shift focus to the highest wind speeds resolved by the model. 
%, which are particularly relevant for applications like wind power forecasting. 
We expect these extreme values to be linked with spatial resolution, as higher-resolution models may better capture more complex atmospheric processes, potentially leading to a more accurate representation of extreme events. In this context, we consider the average of the 100 highest wind speeds to evaluate model performance. Percentile-based evaluation is avoided, as differences in model resolution would result in varying data points for the same percentile across models.

\paragraph{Linear regression}
The CMIP6 datasets differ in their latitude and longitude coordinates, and we use the number of spatial points within the selected area to measure the grid resolution. To examine trends in distance metrics and maximum wind speeds with increasing spatial resolutions, we perform linear-log regression \citep{montgomery2021introduction}. We can then describe the distance of each model to ERA5 as a function of the resolution $r$
\begin{equation}
    w_{max}(r) = \beta_0 + \beta_1\log(r) + \epsilon 
\end{equation}
    
where $\epsilon \sim \mathcal{N}(0, \sigma^2)$ is Gaussian noise and $\beta_0$ and $\beta_1$ are computed using the method of least squares. 
\begin{comment}
\subsubsection{Extreme Wind Speeds}
Due to the non-linear relationship between wind speed and power, different percentiles in the wind speed distribution contribute unequally to wind power generation. Therefore, we highlight distributional differences between GCMs and ERA5 for different wind speed ranges. 

\paragraph{Maximum Values}
In addition to overall distributions, we analyze the maximum wind speeds projected by the models. 

\paragraph{KDE-Ratios}
We investigate high wind speeds by visualizing distributional differences in these critical wind speed ranges using KDE-ratios. We consider distributional differences separately for wind speeds within the critical range defined by the cut-in and cut-off wind speed, given a particular turbine type. Additionally, we consider distributional differences in extreme wind speeds only, which are defined as values above the cut-off wind speed.

\paragraph{Survival Function}
To further visualize the tail behavior of the wind speed distribution, we compute the log of the survival function 

\begin{equation}
    \text{S}(w) = 1 - F_{n}(w) \text{\,,}
\end{equation}

where $F_{n}(w)$ is the empirical cumulative function. 
\end{comment}

\subsubsection{Wind power estimation}
So far, we focused on wind speed, as wind directions are a direct output from GCMs. However, wind power is our variable of interest and its estimation involves assumptions and simplifications. While wind speed is a proxy, the ultimate goal is to understand how variations in wind speed distributions impact (multi-decadal) wind power estimation.
To compute the wind power generation at each grid point, we input the wind speeds at grid points into a wind turbine power curve \citep{sabine_haas_2024_10685057}. In the main part of the paper the chosen wind turbine power curve is the \texttt{Vestas V126-3.45} turbine, with a hub-height of $126 \text{ m}$, see \Cref{fig:power-curve}. We provide supporting results for other turbines in the \Cref{appendix}. 
Finally, we compute the cumulative wind power generation by summing over the power generation of all time steps and grid points. The power predictions from the climate models are normalized by the ERA5 predictions to compare relative cumulative wind power estimates across different models.

\begin{figure}[H]
    \centering    \includegraphics[width=0.5\linewidth]{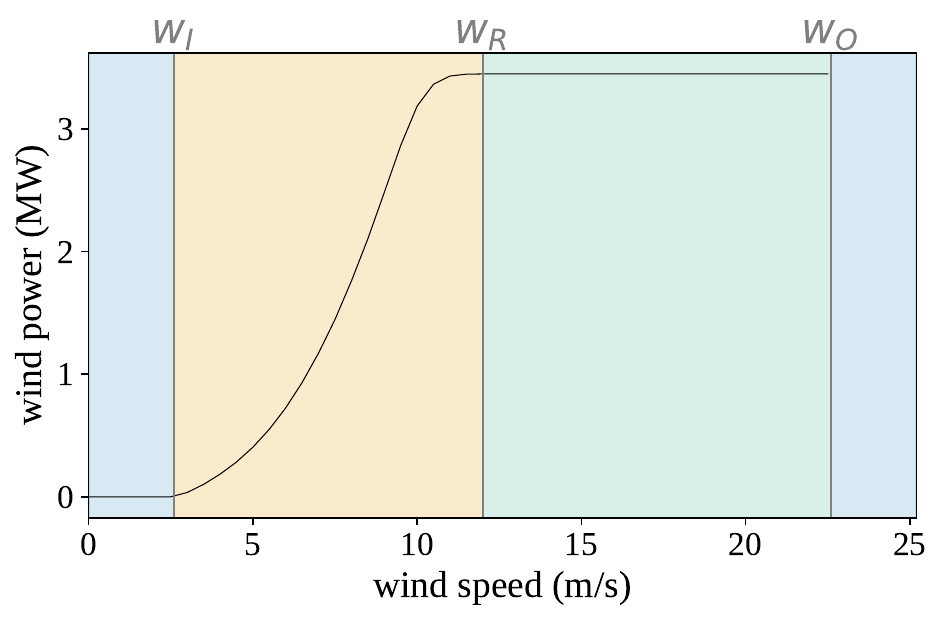}
    \caption{Power curve of the \texttt{Vestas V126-3.45} turbine. No power is generated for wind speeds below the cut-in wind speed $w_i$ and above the cut-out wind speed $w_o$. Between $w_i$ and the rated wind speed $w_r$ the relationship between speed and power is almost cubic. Between $w_r$ and $w_o$ the power output is maximal.}
    \label{fig:power-curve}
\end{figure}

\section{Results}
We investigate the spatial resolution of GCMs and compare wind power predicted by different RCMs. Our results reveal that the choice of the underlying physical model- whether it is an RCM or GCM- is influential, and we do not find systematic trends with increasing resolution.
\label{results}
        \begin{comment}

    \begin{figure}
    \begin{subfigure}[b]{0.48\textwidth}
        \centering
        \includegraphics[width=\textwidth]{fig/jensen-shannon_distance_heatmap_europeland_126m_2005-2015_first-runs.pdf}
    \end{subfigure}
    \hfill
    \begin{subfigure}[b]{0.48\textwidth}
        \centering
        \includegraphics[width=\textwidth]{fig/wasserstein-1_distance_heatmap_europeland_126m_2005-2015_first-runs.pdf}
    \end{subfigure}
    \caption{Heatmap with model rankings by \gls{js} and \gls{w1} distance.}
    \end{figure}
    \end{comment}

\subsection{Higher resolution of GCM does not imply better forecast}
\begin{figure}[H]
    \centering
    \includegraphics[width=\textwidth]{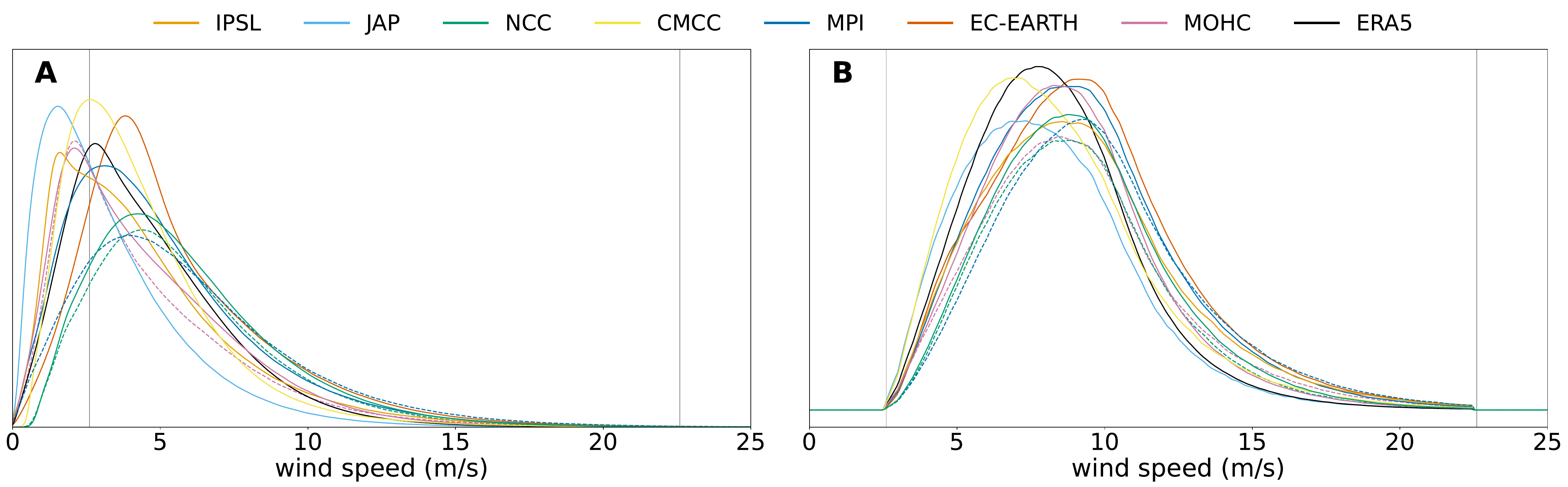}
    \caption{Kernel density estimations of the wind speeds (A) and of the corresponding wind power output (B). When datasets of the same model with different resolutions exist, the low-resolution version is indicated by a dashed line. Horizontal lines indicate the cut-in and cut-out wind speeds of the \texttt{Vestas V126-3.45} turbine considered in this study. While most wind speeds are $< \SI{10}{\meter\per\second}$, most of the wind power is generated by wind speeds $\sim \SI{10}{\meter\per\second}$.}
    \label{fig:wind-distributions}
\end{figure}

\begin{figure}[H]
    \centering
    \begin{subfigure}[b]{\textwidth}
        \centering     
        \includegraphics[width=\linewidth]{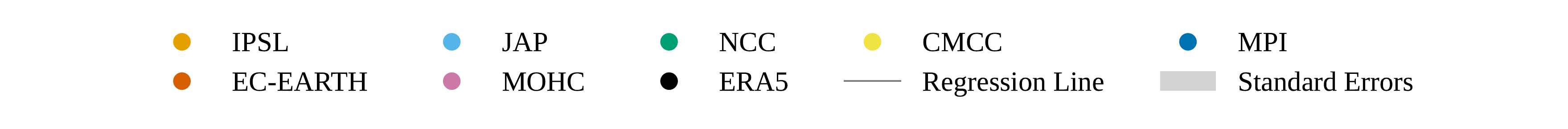}
    \end{subfigure}
    \begin{subfigure}[b]{\textwidth}
        \centering        
        \includegraphics[width=\textwidth]{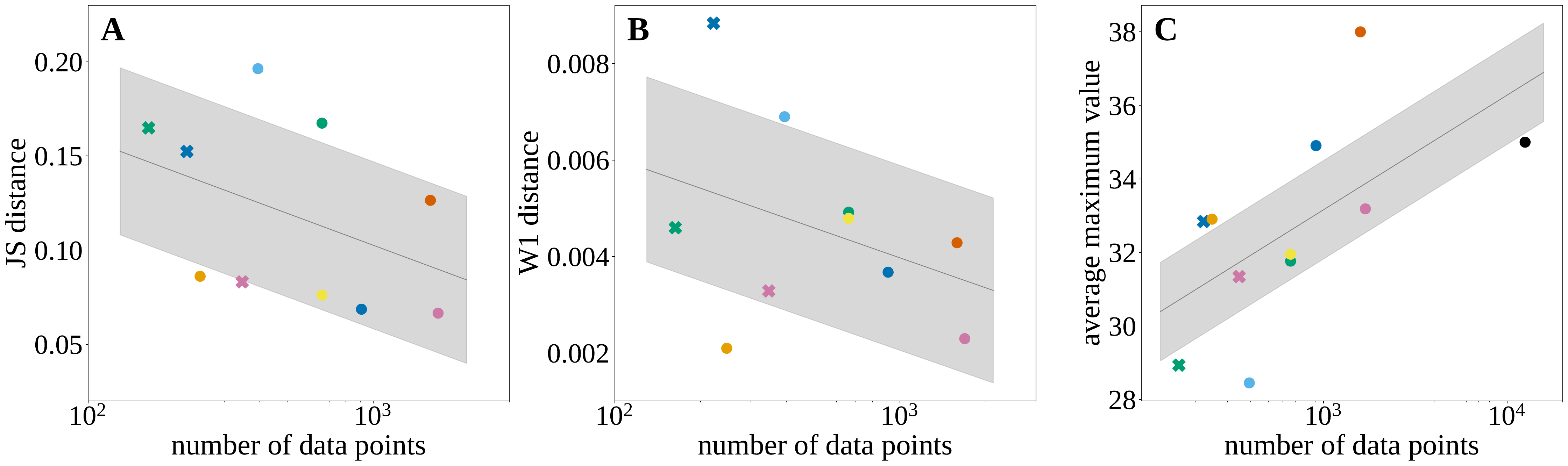}
    \end{subfigure}
\caption{The JS distance (A) and the W1 distance (B) between the wind speeds from CMIP6 and ERA5 and the average over the $100$ highest wind speed values (C) are plotted against the spatial resolution (values are given in \Cref{tab:values_js_w1_max}). When datasets of the same model with different resolutions exist, the low-resolution version is indicated by a cross instead of a circle. The gray line represents the log-linear regression fits with standard errors. The regression in (A) and (B) is not statistically significant at a 5\% significance level, while the regression in (C) is (see \Cref{tab:regression_details}).}
\label{fig:trends}
\end{figure}

The results in \Cref{fig:wind-distributions} reveal substantial differences in the estimated wind speed KDEs across models. For instance, the JAP model notably deviates from the ERA5 wind speed distribution showing a distinct peak in the lower wind speed range. 
We analyze the JS distance and the W1 distance of the GCMs and ERA5 to quantify these distributional differences in wind speeds, as shown in \Cref{fig:trends}. These metrics show substantial variability in each model's ability to capture the wind speed distribution accurately. 
%Although on average both metrics decrease slightly with a higher spatial resolution in a linear regression - indicating a closer match with the ERA5 distribution - this relationship between wind speed accuracy and spatial resolution is not statistically significant. 
However, regression analysis shows no statistically significant relationship between the two metrics considered and spatial resolution.
This suggests that, although the accuracy of wind speed distributions varies largely between models, this variation is not systematically linked to spatial resolution. 

The regression between the average of the $100$ highest wind speeds and the logarithm of spatial resolution is significant at the $5 \%$ level. This suggests a positive relationship between spatial resolution and the model's ability to represent high wind speeds.

When wind speeds are converted to wind power, accurately capturing specific wind speed quantiles becomes critical. Despite this, the results for cumulative wind power estimated for ten years, shown as a percentage relative to ERA5 power output in \Cref{fig:cumulative_cmip6}, reflect similar trends observed in the wind speed distributions. There is considerable variation in the estimated cumulative wind power across different models, with eight models overestimating wind power output relative to ERA5, while two models underestimate it. 

Furthermore, relative cumulative power forecasts differ widely, ranging from $-42.56\%$ to $68.16\%$ when being compared to ERA5. For example, the GCM IPSL shows the closest alignment with ERA5. When considering the GCMs CMCC and MOHC, the forecasts fall within a range that could be useful for practical applications. This raises the question: given that GCMs can already provide valuable forecasts, is the added complexity and computational cost of regional climate models justified?

\begin{comment}
\begin{itemize}
    \item observation: wind speed distributions look different but are Weibull-distributed -> no systematic changes with resolution visible. Same for wind power, some wind speeds become unimportant in power transformation. 
    \item We compare the wind speed distributions using w1 and JS divergence. Results reveal that higher resolution does not necessarily mean that the distribution is closer to reanalysis ERA5. 
    \item Relative cumulative power forecasts are different. They range from x\% to y\%. GCM z is closest to ERA5, if we take GCM z(, a) and b into account, we might be in a useful range.
\end{itemize}
If GCMs give us valubale forecasts, do we even need regional climate model data
\end{comment}
\subsection{RCM choice significantly influences the forecast}
\begin{figure}[H]
 \begin{minipage}[b]{0.49\textwidth}
        \centering
        \includegraphics[width=\linewidth]{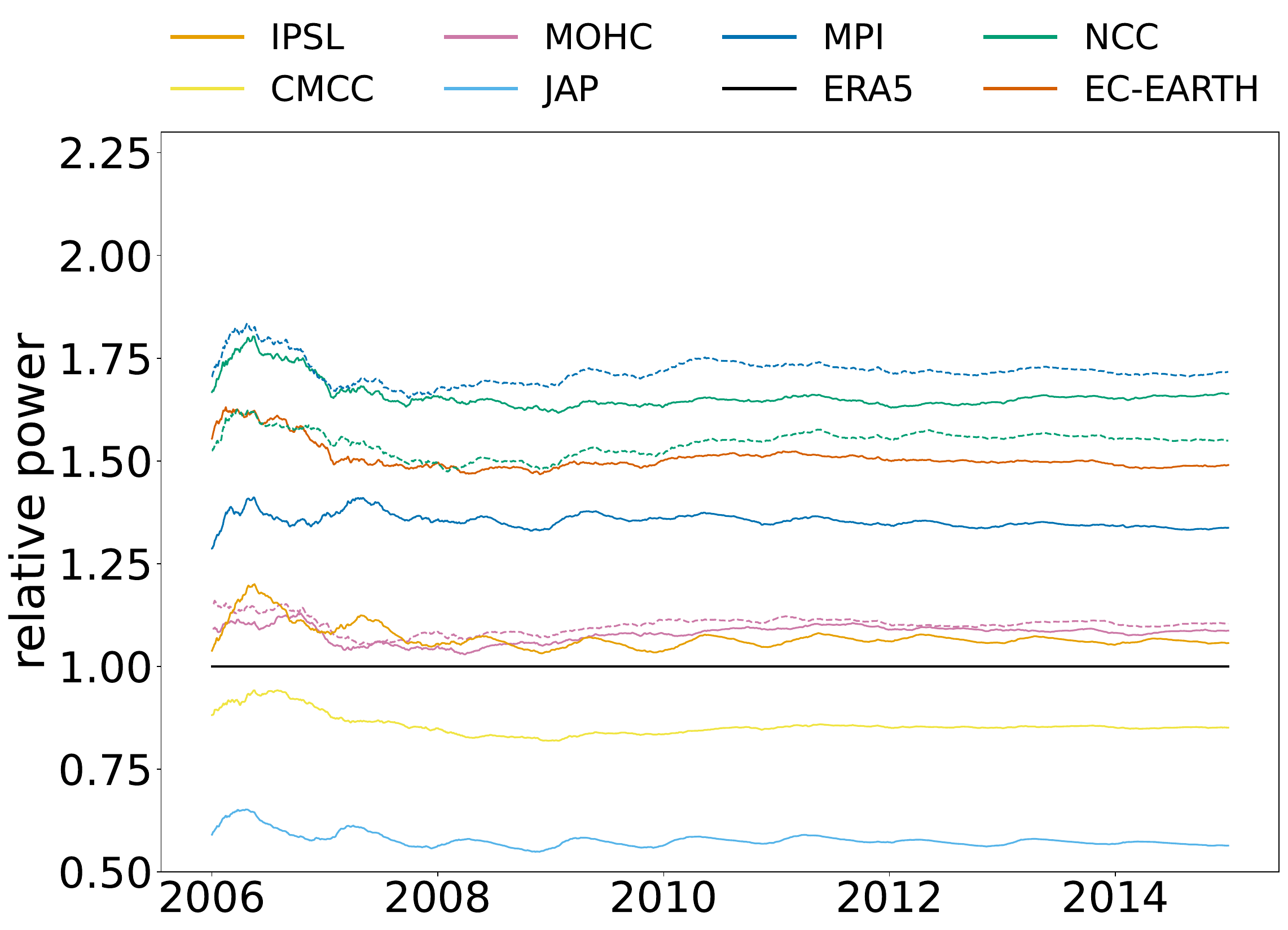}  
        \caption{Relative cumulative power of CMIP6 data sets. When datasets with different resolutions run by the same model exist, the low-resolution version is indicated with a dashed line. The choice of GCM has a substantial influence on the power prediction; most GCMs overestimate wind power relative to ERA5.}
        \label{fig:cumulative_cmip6}
    \end{minipage}
\hspace{0.01\textwidth}
    \begin{minipage}[b]{0.49\textwidth}
        \centering
    \includegraphics[width=\linewidth]{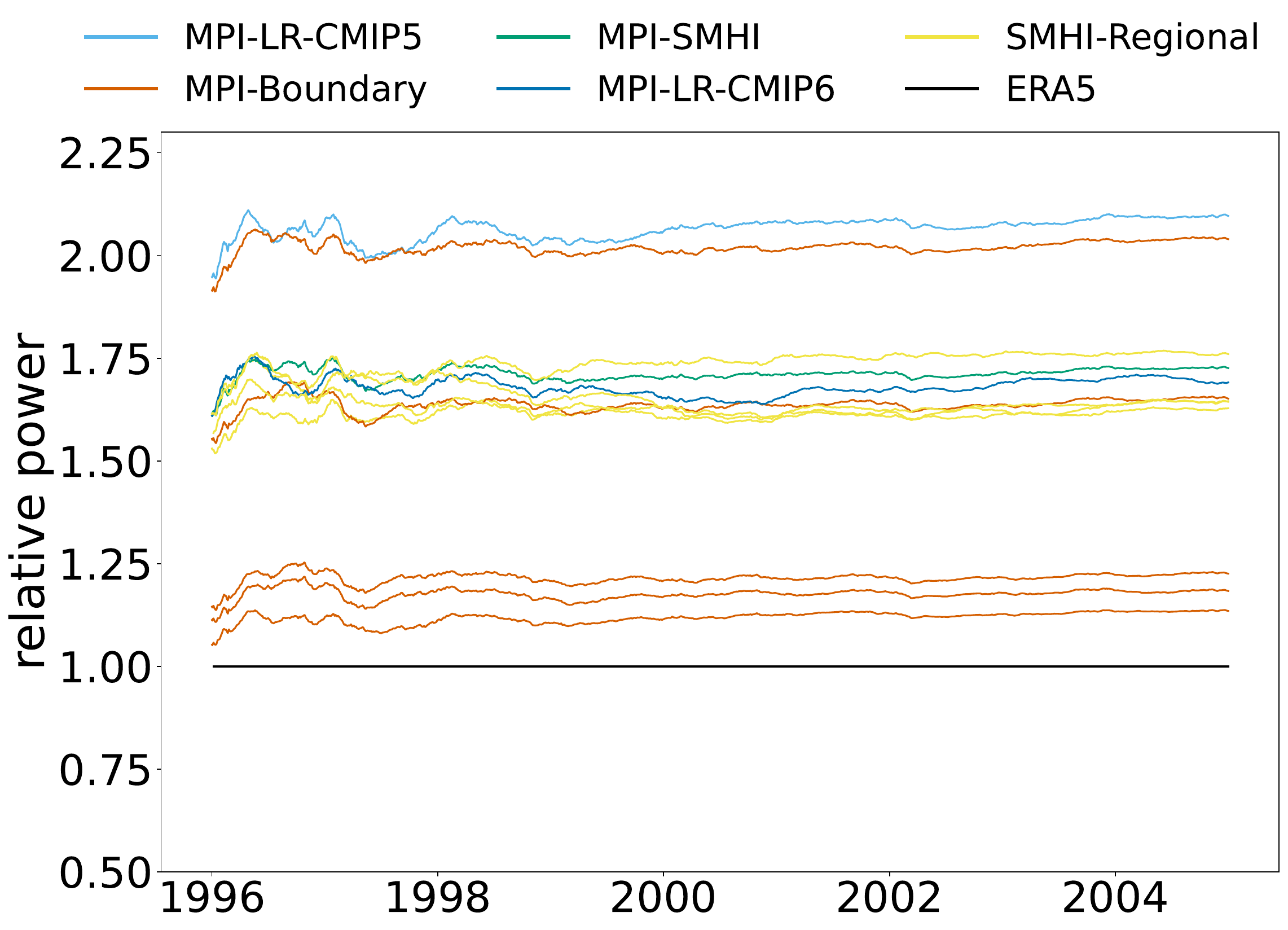}  % Maintain aspect ratio
        \caption{Relative cumulative power of CMIP5 regional and CMIP6 MPI from 1995 to 2005. The spread of the cumulative wind power predictions from multiple RCMs using the output of the MPI model as boundary conditions is larger than the spread using one RCM with different boundary conditions.} 
        \label{fig:cumulative_cmip5}
        \end{minipage}
\end{figure}

We further investigate regional model data from CMIP5 and present the corresponding results in \Cref{fig:cumulative_cmip5}. For the global model run of the MPI CMIP5 model, we use data extrapolated from a different height (see \Cref{methods}). The predictions from CMIP5 overestimate wind power most compared to ERA5. 
In contrast, all regional model runs and the CMIP6 model run are closer to ERA5. Comparing the spread of the predictions from the MPI model that is downscaled with several regional models (orange in \Cref{fig:cumulative_cmip5}) to the prediction of several GCMs downscaled with the regional model SMHI (yellow in \Cref{fig:cumulative_cmip5}) reveals that the spread of different regional climate models is bigger than that of one regional model with different boundary conditions. This is also reflected by the variance of the relative power predictions, which is 10.87\% for the regional models with the same boundary conditions and 0.26\% for the MPI model with different boundary conditions. 

\section{Discussion}
\label{discussion}
% WHAT WE DID (IMPACT SPATIAL RESOLUTION, ANALYSIS, WIND POWER FOCUS)
We evaluate how GCM's spatial data resolution affects the reliability of simulated wind speeds for multi-decadal wind power forecasts. Specifically, we focus on ten raw ensemble members of different CMIP6 GCMs over the period $2005 - 2015$, comparing wind speed simulations and derived wind power estimates for the European region against ERA5 reanalysis data. 
%We additionally investigate whether dynamical downscaling adds value by comparing the performance of several regional models from CMIP5. 
Additionally, we investigate whether regional downscaling adds value by comparing the performance of several regional models from CMIP5. 
We employ various statistical modeling and verification techniques, including directly evaluating wind power to account for the non-linear relationship between wind speed and generated power. 

% MODEL CHOICE VS. SPATIAL RESOLUTION
Contrary to the common assumption that a higher resolution provides more reliable wind speeds \citep{molina2022added, pryor2012influence}, we find no clear trend in wind accuracy with increasing spatial resolution. Instead, we observe that the accuracy of wind speed simulations varies substantially across different models. This suggests that the choice of model has a greater impact on wind speed accuracy than spatial resolution itself. Our analysis of regional climate models shows only a small spread in predictions when using different boundary conditions with the same RCM. A much larger spread is visible when comparing different RCMs, further emphasizing the importance of model choice. Therefore, selecting an appropriate climate model should be approached case-by-case, guided by a systematic procedure like the one we present in this paper. 

%Our analysis of regional climate model data supports this claim as we find that the choice of regional model influences the predictions more than the GCM boundary conditions. % This implies that a diverse selection of RCMs may be more valuable than a large ensemble of GCMs.  
%Our results indicate that a diverse selection of RCMs--no matter the underlying GCM-- may provide a useful spread of simulations, offering valuable insights for applications that require a reliable assessment of uncertainty. 

% IMPLICATIONS OF MAIN FINDINGS
The accuracy of GCMs depends on their ability to capture key climate processes, forcings, and feedback mechanisms \citep{doblas2021linking}. % maybe cite: (IPCC; Chapter 10) 
Increasing spatial resolution alone does not guarantee an improved representation of these complex processes. This aligns with findings on effective resolution \citep{klaver2020effective}, where increases in spatial resolution do not proportionally translate into enhanced representation of meteorological phenomena, as measured by global kinetic energy spectra. Therefore, we advocate for using CMIP6 models for multi-decadal wind power forecasts. Although CMIP6 may have a coarser resolution, it offers notable advancements over its predecessor CMIP5 in simulating climate dynamics and providing more reliable multi-decadal forecasts \citep{miao2023evaluation}. 
In line with this rationale, we refrain from applying bias correction or downscaling in our analysis, recognizing that these techniques cannot correct fundamental model misrepresentations and may even disrupt the internal consistency of the climate models \citep{maraun2016bias, franccois2020multivariate}. 

% BETTER WIND SPEED TAILS
Although our analysis shows no clear trend in wind speed accuracy with increasing spatial resolution, we do find that high-resolution models consistently perform better in capturing the upper tail of the wind speed distribution. This suggests that the benefits of (high-quality) high-resolution models are particularly critical in applications focused on extreme wind events, such as risk assessment. 

% BEST MODEL & LIMITATIONS
The IPSL model performs exceptionally well in simulating wind speeds and wind power over Europe, while the MPI-LR model performs the least well. 
%The difference in performance possibly stems from their different focuses in climate system modeling. 
However, our findings are likely region-specific and several limitations must be considered. We simplify our analysis by modeling wind power with a theoretical power curve and interpolating vertically to model wind speeds at a standard hub height. Although ERA5 reanalysis is widely recognized as a reasonable reference \citep{molina2021comparison, olauson2018era5, kaspar2020regional}, it still contains notable biases that affect wind power simulations \citep{staffell2016using}.

\section{Conclusion}
Climate information is essential for multi-decadal wind power forecasting. Available climate models, among other things, vary substantially in their spatial resolution. We analyze 11 global and regional climate models for their ability to predict long-term wind power. 

Our findings highlight the need for high-quality GCMs and RCMs, particularly for wind energy planning. Given the sensitivity of wind power generation to local climate conditions, it is essential that models accurately capture both large-scale dynamics and more local meteorological phenomena. Higher-resolution climate models enhance the representation of extreme wind speeds, yet they do not guarantee improved accuracy in wind speed forecasts. Instead, the selection of the climate models, both GCM and RCM, emerges as a more influential factor for wind speed and power predictions than the spatial resolution itself. As we do not find substantial differences between CMIP5 regional models and CMIP6 global models, we recommend selecting climate models from CMIP6 for multi-decadal wind speed forecasts, as they offer the most current advancements in climate modelling.
%Regional models can add value in specific contexts, which was shown in previous works. 
%The analysis shows that GCMs are useful for wind power forecasting, and RCMs add further value, emphasizing the need for high-quality models at both global and regional scales.

\section*{Acknowledgements}
This research was funded by the Deutsche Forschungsgemeinschaft (DFG, German Research Foundation) under Germany’s Excellence Strategy – EXC number 2064/1 – Project number 390727645, the Tübingen AI Center, and the Athene Grant of the University of Tübingen. The authors thank the International Max Planck Research School for Intelligent Systems (IMPRS-IS) for supporting Nina Effenberger and Luca Schmidt. 

\bibliographystyle{plainnat}
\bibliography{bib}
\newpage
\newpage
\begin{appendices}

\counterwithin{figure}{section}
\counterwithin{table}{section}
\section{Supplementary Material}
\label{appendix}

\begin{table}[htp]
    \centering
    \caption{CMIP6 data sets: Labels as used in this paper (first column), names of atmospheric model components (second column), original spatial resolutions in terms of grid specification (third column), number of data points when selecting the land mass of Europe (fourth column), and native resolution (fifth column) as indicated in the references given in the last column.}
    \label{tab:models_information}
    \begin{tabular}{ l | l l l l l}
    \toprule
        Label & Model  & Grid  & Number of & Native & Reference \\
         & & Resolution & Data Points & Resol. &  \\
    \midrule
        CMCC & CM2-SR5 & 0.9° × 1.25° & 661 &  100 km & \citet{CMCC} \\
        EC-EARTH & EC-Earth3.3 & T255
        %linearly reduced Gaussian grid 
        & 1586 & 100 km & \citet{EC-Earth} \\
        IPSL & CM6A-LR & 2.5° × 1.3°
        % N96
        & 247 & 250 km & \citet{IPSL} \\
        JAP & MIROC6 &  T85 & 394 & 250 km & \citet{JAP}\\
        MOHC-LR & HadGEM3-GC31-LL & N96
        %(135 km in midlatitudes) 
        & 347 & 250 km & \citet{MOHC-LR} \\
        MOHC-HR & HadGEM3-GC31-ML & N216
        %(60 km  in midlatitudes) 
        & 1688 & 100 km & \citet{MOHC-HR} \\
        MPI-LR & ESM1.2-LR & T63 & 222 & 250 km & \citet{MPI-LR} \\
        MPI-HR & ESM1.2-HR & T127 & 909 & 100 km & \citet{MPI-HR} \\
        NCC-LR  & NorESM2-LM &  2.5° × 1.875° & 163 & 250 km & \citet{NCC-LR} \\
        NCC-HR  & NorESM2-MM & 1.25° × 0.9375° & 661 & 100km & \citet{NCC-HR} \\
        MPI-LR-CMIP5 & ECHAM6 & T63 & 222 & 200 km & \citet{MPI_CMIP5_historical} \\
        ERA5 & IFS CY41R2 & 0.25° × 0.25° & 12506 &  & \citet{era5_data} \\ %T639 (spectral)
    \bottomrule
    \end{tabular}
\end{table}

\begin{table}[htp]
    \centering
    \caption{CORDEX data sets: Labels of the global boundary models (first column) and their full model names (second column), labels of the regional models (third column) and their full model names (fourth column).  All data sets have resolution 0.11° × 0.11° ~12.5km and were withdrawn from \citet{CORDEX_2019}.}
    \label{tab:models_information_rcms}
    \begin{tabular}{ l l l l }
    \toprule
        Label & Model & Label & Model \\
        GCM & & RCM & \\
    \midrule
        MPI & MPI-M-MPI-ESM-LR & CNRM & CNRM-ALADIN63 \\
        MPI & MPI-M-MPI-ESM-LR & DMI & DMI-HIRHAM5 \\
        MPI & MPI-M-MPI-ESM-LR & ETH & CLMcom-ETH-COSMO-crCLIM \\
        MPI & MPI-M-MPI-ESM-LR & ICTP & ICTP-RegCM4-6 \\
        MPI & MPI-M-MPI-ESM-LR & MOHC & MOHC-HadREM3-GA7-05 \\
        MPI & MPI-M-MPI-ESM-LR & SMHI & SMHI-RCA4 \\
        MOHC & MOHC-HadGEM2-ES & SMHI & SMHI-RCA4 \\
        NCC & NCC-NorESM1-M & SMHI & SMHI-RCA4 \\
        IPSL & IPSL-CM5A-MR & SMHI & SMHI-RCA4 \\
        EC-EARTH & ICHEC-EC-EARTH & SMHI & SMHI-RCA4 \\
        CNRM & CNRM-CERFACS-CM5 & SMHI & SMHI-RCA4 \\
    \bottomrule
    \end{tabular}
\end{table}

\begin{table}[H]
\caption{Characteristics of the wind speed samples for all considered models: the mean, the average of $100$ highest wind speed values, the Jensen-Shannon (JS), and the Wasserstein-1 (W1) distance to ERA5.}
\label{tab:values_js_w1_max}
\centering
\begin{tabular}{ l|lllllll }
\toprule
Model & Mean & Max average & JS distance & W1 distance \\
\midrule
ERA5 & 4.58 & 35.00 & - & - \\
NCC-LR & 5.80 & 28.94 & 0.165 & 0.005 \\
NCC-HR & 5.84 & 31.77 & 0.167 & 0.005 \\
MPI-LR & 5.92 & 32.84 & 0.152 & 0.009 \\
MPI-HR & 5.05 & 34.91 & 0.069 & 0.004 \\
MOHC-LR & 4.50 & 31.34 & 0.083 & 0.003 \\
MOHC-HR & 4.55 & 33.19 & 0.067 & 0.002 \\
JAP & 3.37 & 28.45 & 0.196 & 0.007 \\
IPSL & 4.56 & 32.91 & 0.086 & 0.002 \\
EC-EARTH & 5.54 & 38.00 & 0.126 & 0.004 \\
CMCC & 4.32 & 31.95 & 0.076 & 0.005 \\
\bottomrule
\end{tabular}
\end{table}

\begin{table}[H]
\caption{Regression details for the $100$ highest wind speed values, the Jensen-Shannon (JS) and Wasserstein-1 (W1) distance including the regression parameters $\alpha$ and $\beta$, the standard error of the slope, the explained variance ($100\cdot R^2$) in \% and the $p$-value of the Overall-$F$-Test.}
\label{tab:regression_details}
\centering
\begin{tabular}{l|lllllllllllll}
\toprule
& Intercept $\alpha$ & Slope $\beta$ & Std. Dev. & $ R^2$ & $p$-value \\
\midrule
JS distance & 0.271 & -0.056 & 0.044 & 16.698 & 0.241 \\
W1 distance & 0.010 & -0.002 & 0.002 & 12.647 & 0.313 \\
Max Average & 23.807 & 3.119 & 1.339 & 37.626 & 0.045 \\
\bottomrule
\end{tabular}
\end{table}

\begin{figure}[H]
    \centering
    \begin{subfigure}[b]{\textwidth}
        \centering     \includegraphics[width=\linewidth]{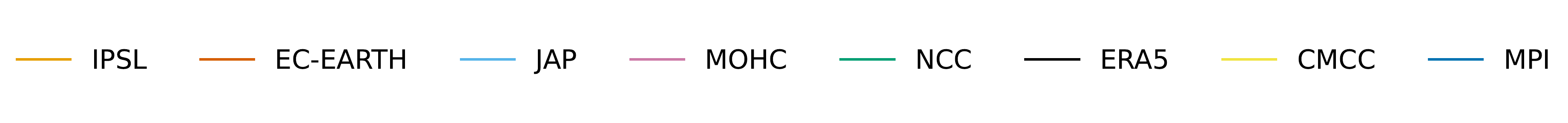}
    \end{subfigure}
    \begin{subfigure}[b]{\textwidth}
        \centering        \includegraphics[width=\textwidth]{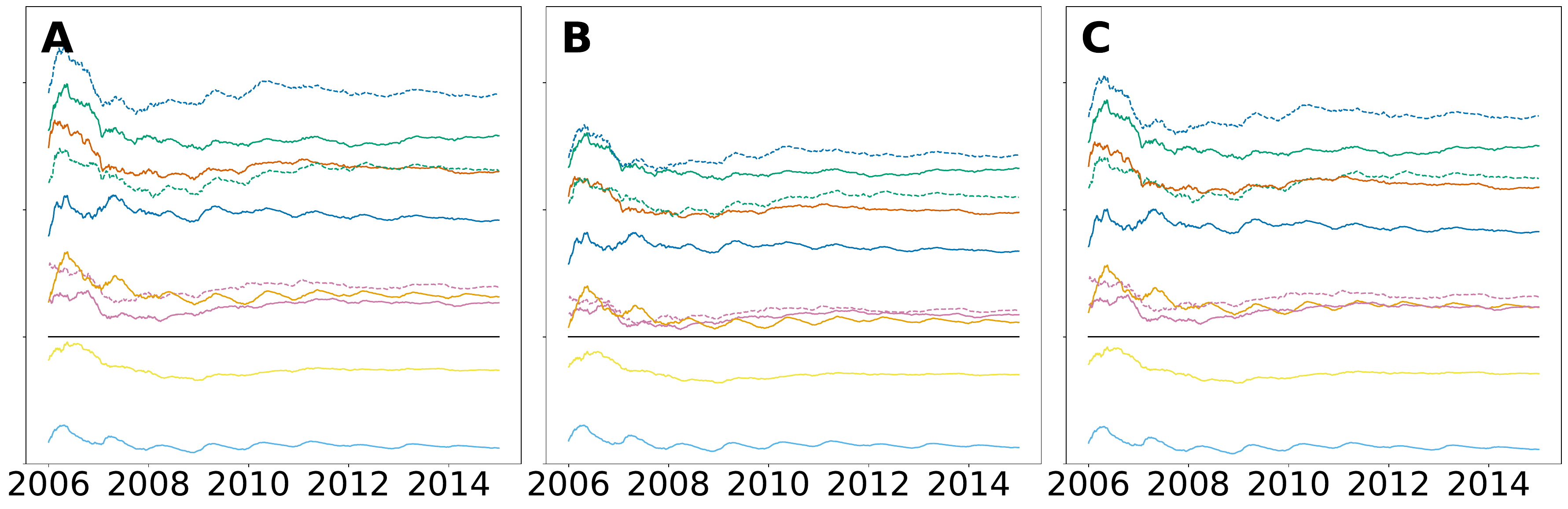}
    \end{subfigure}
\caption{Relative cumulative wind power for turbine \texttt{Enercon E-70/2.00} at hub height \SI{85}{\meter} (A), turbine \texttt{Vestas V126-3.45} at hub height \SI{126}{\meter} like in the main text (B) and turbine \texttt{Vestas V164/9.50} at hub height \SI{140}{\meter} (C).}
\label{fig:cumulative_wp_turbine_comparison}
\end{figure}

\end{appendices}
\end{document}